\documentclass[twocolumn,amsmath,amssymb,superscriptaddress,prl]{revtex4}%
\usepackage{epsf}
\usepackage{epsfig}
\usepackage{graphicx}
\usepackage{dcolumn}
\usepackage{bm}
\usepackage{amsmath}
\usepackage{amsfonts}
\usepackage{amssymb}%
\providecommand{\U}[1]{\protect\rule{.1in}{.1in}}
\input epsf
\begin{document}
\title{Controlling the conductance and noise of driven carbon-based Fabry-P\'{e}rot devices}
\author{Luis E. F. Foa Torres}
\affiliation{Institute for Materials Science and Max Bergmann Center of Biomaterials,
Dresden University of Technology, D-01062 Dresden, Germany}
\author{Gianaurelio Cuniberti}
\affiliation{Institute for Materials Science and Max Bergmann Center of Biomaterials,
Dresden University of Technology, D-01062 Dresden, Germany}
\date{\today}

\begin{abstract}
We report on ac transport through carbon nanotube Fabry-Perot devices. We show
that tuning the intensity of the ac gating induces an alternation of
suppression and partial revival of the conductance interference pattern. For
frequencies matching integer multiples of the level spacing of the system
$\Delta$ the conductance remains irresponsive to the external field. In
contrast, the noise in the low bias voltage limit behaves as in the static
case only when the frequency matches an \textit{even} multiple of the level
spacing, thereby highlighting its phase sensitivity in a manifestation of the
\textit{wagon-wheel effect in the quantum domain}.

\end{abstract}

\maketitle



Achieving control of the electrical response of nanometer scale devices by
means of external fields is a main goal of nanoelectronics. This quest for
control is usually pursued using static electric or magnetic fields, whereas
the use of time-dependent excitations \cite{Platero2004,Kohler2005} remains
much less explored. Besides offering captivating phenomena
\cite{Thouless1983,Grossmann1991,Wagner,Altshuler1999}, such as the generation
of a dc current at zero bias voltage
\cite{Thouless1983,Altshuler1999,Pumping-recent}, understanding the influence
of ac fields becomes necessary if nanoscale devices are going to be integrated
in everyday electronics.

Carbon based materials including fullerenes, graphitic systems and polymers,
are promising building blocks for these devices. Among them, carbon nanotubes
\cite{Charlier2007} stand out due to their extraordinary mechanical and
electrical properties \cite{Charlier2007}. Applications include transistors,
sensors and interconnects \cite{Coiffic2007}. As compared to other molecular
systems, single walled carbon nanotubes offer the unique opportunity of
achieving almost ballistic transport. Indeed, thanks to the low resistance of
the nanotube-electrodes contacts, Fabry-Perot (FP) conductance oscillations
were experimentally observed \cite{Liang2001} and the current noise measured
\cite{Wu2007}. Although some experimental \cite{Kim2004} and theoretical
\cite{Roland2000,Orellana2007,Wuerstle2007,Guigou2007,Oroszlany2009} studies
on the effects of ac fields on nanotubes are available, the effect of ac
fields on the conductance and noise in this Fabry-Perot regime is unknown: Is
it possible to tune the interference pattern through ac fields? Can ac fields
help us to unveil phase sensitive information encoded in the current noise?

\begin{figure}[ptbh]
\includegraphics[width=8.0cm]{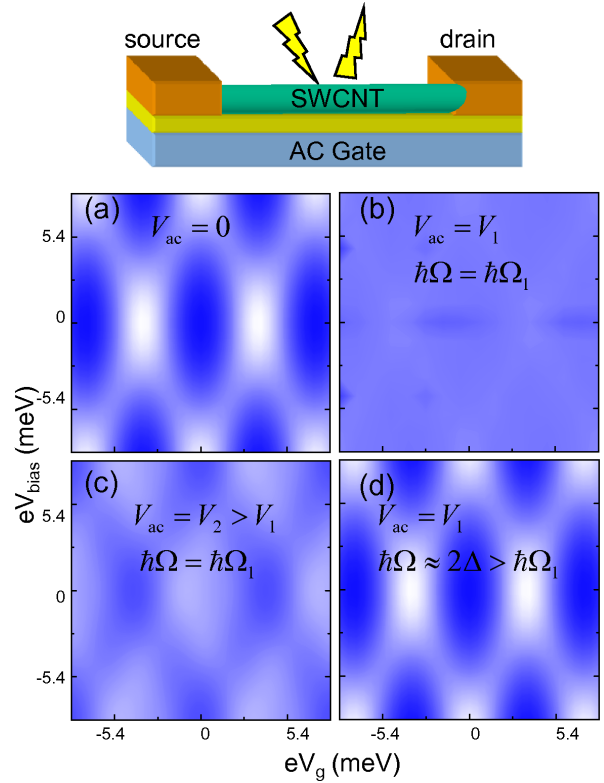} \caption{(color online) Top, scheme of
the device considered in the text, a CNT connected to two electrodes and an ac
gate. The panels marked with a, b, c and d, are the Fabry-Perot conductance
interference patterns (as a function of bias and gate voltages) observed with
no ac gating (panel a) is modified when different different driving
frequencies and amplitudes are applied: b (suppression), c (revival and phase
inversion) and d (robustness). White and dark blue correspond to maximum and
minimum conductances respectively.}%
\label{fig1}%
\end{figure}

In this Letter, we answer the above questions by modelling the electronic
transport through driven carbon nanotube FP resonators. We show that by tuning
the frequency and intensity of a harmonic gate (see scheme in Fig. \ref{fig1})
one can control not only the conductance but also the current noise
\cite{Blanter2000}. Moreover, we show that the current noise carries phase
sensitive information not available in the static case.

\textit{Tight-binding model and Floquet solution. }Several approaches can be
used to describe time-dependent transport. They include: the Keldysh formalism
\cite{Pastawski1992,Jauho1994,Arrachea2005b}, schemes that use density
functional theory \cite{Kurth2005}, the equation of motion method
\cite{Agarwal2007}, and schemes that exploit the time-periodicity of the
Hamiltonian through Floquet theory \cite{Moskalets2002,Camalet2003}. Here, as
a general framework we use the Floquet scheme \cite{Camalet2003} combined with
the use of Floquet-Green functions \cite{FoaTorres2005}. Within this
formalism, the dc component of the time dependent current as well as the dc
conductance (called simply conductance hereafter), can be fully written in
terms of the Green's functions for the system \cite{Kohler2005,FoaTorres2005}.
The current noise can be obtained from the correlation function $S(t,t^{\prime
})=\left\langle \left[  \Delta I(t)\Delta I(t^{\prime})+\Delta I(t^{\prime
})\Delta I(t)\right]  \right\rangle ,$ $\Delta I(t)=I(t)-\left\langle
I(t)\right\rangle $ being the current fluctuation operator. The noise strength
can be characterized by the zero frequency component of this correlation
function averaged over a driving period, $\bar{S},$ which can be casted in a
convenient way within this formalism \cite{Kohler2005}. Further
simplifications can be achieved by using the broad-band approximation and an
homogenous gating of the tube \cite{Kohler2005}.

For simplicity we consider an infinite CNT described through a $\pi$-orbitals
Hamiltonian \cite{Charlier2007} $H_{e}=\sum_{i}E_{i}^{{}}c_{i}^{+}c_{i}^{{}%
}-\sum_{\left\langle i,j\right\rangle }[\gamma_{i,j}c_{i}^{+}c_{j}^{{}%
}+\mathrm{H.c.}]$, where $c_{i}^{+}$ and $c_{i}^{{}}$ are the creation and
annihilation operators for electrons at site $i$, $E_{i}$ are the site
energies and $\gamma_{i,j}$ are nearest-neighbors carbon-carbon hoppings. To
model the FP interferometer, a central part of length $L$ (the
\textquotedblleft sample\textquotedblright) is connected to the rest of the
tube through matrix elements $\gamma_{t}$ smaller than the hoppings in the
rest of the tube which are taken to be equal to $\gamma_{0}=2.7$~$\mathrm{eV}$
\ \cite{Charlier2007}.\ $L$ can be used to tune the level spacing
$\Delta\propto1/L$. For the case of uniform gating of the sample, it is
modeled as an additional on site energy $E_{j\in\mathrm{CNT}}=eV_{g}%
+eV_{\mathrm{ac}}\cos(\Omega t)$. This non-interacting model is justified when
screening by a metallic substrate or by the surrounding gate lessens
electron-electron interactions. When these interactions come into play effects
beyond our present scope may emerge \cite{Guigou2007}.\begin{figure}[ptb]
\includegraphics[width=7.0cm]{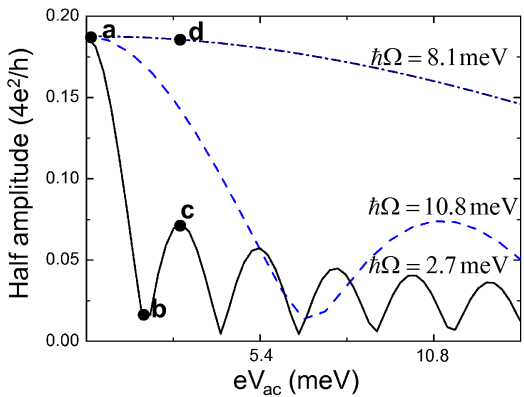} \caption{(color online) Half amplitude
of the FP conductance oscillations as a function of the ac field intensity.
These results are for a metallic zig-zag nanotube of length $L=440$%
~$\mathrm{nm}$ and $\gamma_{t}=0.7\gamma_{0}$, at zero bias voltage and
temperature.}%
\label{fig2}%
\end{figure}

\textit{Suppression, revival and robustness of FP oscillations under AC
gating. }In the following we consider a uniform gating of the tube, the same
effects are expected for an ac bias or illumination with radiation of
wavelength larger than the device length as in Ref. \cite{Keay1995} where
frequencies of up to $3$THz were achieved. At low to moderate frequencies
($\hbar\Omega<\Delta$), our main observation is that the amplitude of the FP
conductance oscillations is reduced and can even be suppressed by tuning the
intensity of the field. This is clearly shown in Fig. \ref{fig2} (solid line)
where the half amplitude of the conductance oscillations (computed at zero
temperature) is shown as a function of the ac field intensity $V_{\mathrm{ac}%
}$. Different curves correspond to different frequencies. Interestingly, for
certain frequencies, the FP pattern is completely suppressed by tuning the
intensity of the ac field, an effect which survives in the adiabatic limit.

\begin{figure}[ptb]
\includegraphics[width=8.5cm]{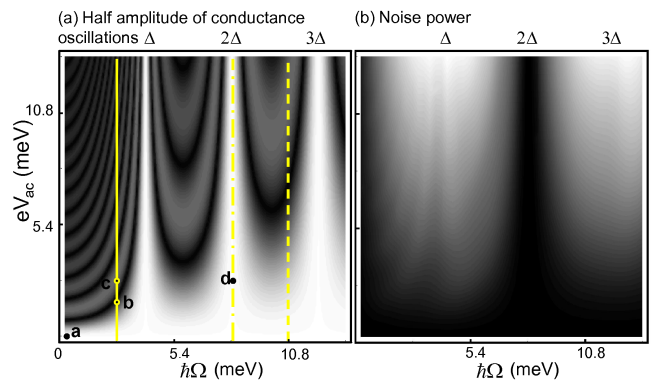} \caption{(color online) a) Contour
plot showing the half amplitude of the FP oscillations as a function of
$V_{ac}$ and $\Omega$ at zero temperature and bias voltage. b) Same contour
plot for the current noise $\bar{S}$. White in both figures is for maximum
values of the half amplitude and noise respectively.}%
\label{fig3}%
\end{figure}

Figure \ref{fig1}, shows the FP\ patterns obtained at the points marked on the
curves in Fig. \ref{fig2} . There, by comparison with the static case (a), one
can observe the suppression (b) and subsequent revival with a phase inversion
(c) of the FP oscillations as $V_{\mathrm{ac}}$ increases. On the other hand,
panel d shows a situation in which the frequency is tuned to meet the
\textit{wagon-wheel} or \textit{stroboscopic condition} ($\hbar\Omega\approx
n\Delta,$ $n$ integer). The overall dependence of the FP amplitude on both
$\hbar\Omega$ and $V_{\mathrm{ac}}$ is better captured by the contour plot in
Fig.~\ref{fig3}-a. White corresponds to maximum values of the half amplitude
and black to zero values. In general, although the field produces no
appreciable change in the conductance whenever $\hbar\Omega\approx n\Delta$, a
static behavior in the noise requires a more stringent condition as we will
see later.

To rationalize these features let us first consider the adiabatic limit. In
this limit, the period of the ac oscillation is long enough such that at each
instant of time the system can be considered as static with an applied field
which coincides with its instantaneous value. Within this approximation, the
conductance is given by $G_{\mathrm{ad}}=G_{\mathrm{avg}}+A\times J_{0}(2\pi
eV_{\mathrm{ac}}/\Delta)\times\cos(2\pi eV_{g}/\Delta)$, where $A$ and
$G_{\mathrm{avg}}$ are the half amplitude and average value of the conductance
in the static situation (vanishing ac field). We can see that the amplitude is
modulated by a factor $J_{0}(2\pi eV_{\mathrm{ac}}/\Delta)$. Thus, the FP
interference is destroyed whenever $2\pi eV_{\mathrm{ac}}/\Delta$ is a root of
$J_{0}$. This relation would allow the experimental determination of the
effective amplitude of the ac gating.

Deviations from the adiabatic result are due when the condition $\hbar
\Omega\ll\Delta$ is not fulfilled. In this case, by assuming the broad-band
approximation, the dc conductance can be written in the Tien-Gordon
\cite{Tien-Gordon} form: $G(\varepsilon_{F},V_{g})=\sum_{n}\left\vert
a_{n}\right\vert ^{2}$ $G_{\mathrm{static}}(\varepsilon_{F}+n\hbar\Omega
,V_{g}),$ where $a_{n}=J_{n}(eV_{\mathrm{ac}}/\hbar\Omega)$ represents the
probability amplitude of emitting (absorbing) $n$ photons. Thus, the averaging
of the conductance takes place only among energies differing in discrete steps
thereby rendering the smoothing of the oscillations less effective. A higher
intensity of the field is therefore needed to suppress the conductance
oscillations. Indeed, whenever $\hbar\Omega$ is commensurate with $\Delta$ the
contrast of the FP pattern is insensitive to the ac field intensity. The
dash-dotted line in Fig. \ref{fig2} shows the amplitude vs. $V_{\mathrm{ac}}$
for a frequency close to such a condition.

Length dependence: In the infinite tube limit ($\Delta\rightarrow0$) the FP
oscillations disappear and, as long as higher subbands do not contribute to
transport, the conductance becomes insensitive to the ac field
\cite{Orellana2007}. For Fermi energies away from the charge neutrality point,
our results remain valid provided that the energy range effectively probed by
the field ($\sim\max(eV_{\mathrm{ac}},N\hbar\Omega)$, where $N$ is the typical
number of photons) is small enough ($\max(eV_{\mathrm{ac}},N\hbar\Omega
)\ll\varepsilon_{F}$), thereby giving generality to our results for systems
other than nanotubes. The reservoirs are assumed to be in thermal equilibrium
and current-induced heating is assumed not to compromise the tube stability.
The typical current-induced heat flow is of about $50$nW, much below the
experimental limits reported in \cite{Shi2009}.

\textit{AC effects on the current noise and quantum wagon-wheel effect. }The
current noise as characterized by the dimensionless Fano factor (not shown
here) is affected by the ac field only at high frequencies and driving
amplitudes ($\hbar\Omega,V_{\mathrm{ac}}\sim\Delta$). More interesting is the
low bias, low temperature limit. For a static Hamiltonian, the noise power
$\overline{S}$ vanishes in this limit as no fluctuations remain in the
electron distributions. This is in striking contrast with the situation in
which an ac field is applied to the conductor where there is always a non-zero
current noise. This is because even at zero temperature and bias, the ac field
introduces probabilistic scattering processes (photon-assisted transitions)
which add uncertainty to the effective electron distributions, thereby giving
a non-vanishing contribution to the current noise \cite{Kohler2005}. In Fig.
\ref{fig3}-b a contour plot of the current noise $\overline{S}$ as a function
of the driving frequency and ac intensity is presented. The color scale ranges
from $6\times10^{-27}$ $\mathrm{A}^{2}/\mathrm{Hz}$ (black) to $2\times
10^{-25}$ $\mathrm{A}^{2}/\mathrm{Hz}$ (white). The thermal contribution to
the noise at $800$mK is estimated to be $6\times10^{-27}$ $\mathrm{A}%
^{2}/\mathrm{Hz}$ and would therefore not be noticeable in this color scale.

Figure \ref{fig3}-b shows that, in contrast to what is observed for the
conductance, whenever the wagon-wheel or stroboscopic condition is attained
$\overline{S}$ does not behave as in the static case (i. e. different from
zero). Indeed, there is a suppression of $\overline{S}$ only for $\hbar\Omega$
commensurate with \textit{twice} $\Delta$. This is due to the fact that the
noise under ac conditions is sensitive to the phase of the transmission
amplitude which changes only by $\pi$ (and not $2\pi$) from one resonance to
the next one. In between these minima there are local maxima whose intensity
is proportional to $V_{\mathrm{ac}}.$ A similar situation was reported for a
double quantum dot \cite{Strass2005}, and for a driven system composed of two
barriers of varying strength and a uniform varying potential in between
\cite{Moskalets2008}. The noise suppression observed here can be understood by
using a simplified expression for the noise as done before for the
conductance. The main conclusion is that when $\hbar\Omega=2n\Delta$
($n$\textit{ }integer), the noise behaves as in the static case. This effect
can be visualized as a\textit{ manifestation of the wagon-wheel effect in the
quantum domain }where a static behavior in the phase sensitive noise requires
a doubling of the stroboscopic frequency.

\textit{Conclusions.} In summary, we have analyzed the effects of ac gating on
the conductance and noise of FP nanotube-based resonators. It was shown that
the ac field can be used to tune the conductance and noise of the device.
Suppression of the FP conductance oscillations can be achieved even in the
adiabatic limit by tuning the driving amplitude. In contrast, when the driving
frequency matches (a multiple of) the level spacing (wagon-wheel condition),
the conductance coincides with the one of the static system. In contrast, the
noise coincides with the static one only when $\hbar\Omega$ is commensurate
with \textit{twice} $\Delta$ (quantum wagon-wheel condition), therefore
highlighting its phase sensitivity. Although here we considered only
nanotubes, our main results are expected to be valid for more general FP resonators.

We thank S. Kohler and M. Moskalets for useful comments and M. del Valle for
discussions. This work was supported by the Alexander von Humboldt Foundation
and by the EU project CARDEQ under contract No. IST-021285-2. Computing time
was provided by ZIH-TUD.


\begin{thebibliography}{99}                                                                                               %


\bibitem {Platero2004}G. Platero and R. Aguado, Phys. Rep. \textbf{395}, 1 (2004).

\bibitem {Kohler2005}S. Kohler, J. Lehmann, and P. H\"{a}nggi, Phys. Rep.
\textbf{406}, 379 (2005).

\bibitem {Thouless1983}D.~J. Thouless, Phys. Rev. B \textbf{27}, 6083 (1983).



\bibitem {Grossmann1991}F. Grossmann, T. Dittrich, P. Jung, and P. H\"{a}nggi,
Phys. Rev. Lett. \textbf{67}, 516 (1991).

\bibitem {Wagner}M. Wagner, Phys. Rev. A \textbf{51}, 798 (1995).

\bibitem {Altshuler1999}B. L. Altshuler and L. I. Glazman, Science
\textbf{283}, 1864 (1999); M. Switkes, C. M. Marcus, K. Campman, and A. C.
Gossard, Science \textbf{283},1905 (1999).

\bibitem {Pumping-recent}B. Kaestner, V. Kashcheyevs, S. Amakawa, M. D.
Blumenthal, L. Li, T. J. B. Janssen, G. Hein, K. Pierz, T. Weimann, U.
Siegner, and H. W. Schumacher, Phys. Rev. B \textbf{77}, 153301 (2008); A.
Fujiwara, K. Nishiguchi, Y. Ono, Appl. Phys. Lett. \textbf{92}, 042102 (2008);
B. Kaestner, C. Leicht, V. Kashcheyevs, K. Pierz, U. Siegner, and H. W.
Schumacher, \textit{ibid.} \textbf{94}, 012106 (2009).

\bibitem {Charlier2007}J.-C. Charlier, X. Blase, and S. Roche, Rev. Mod. Phys.
\textbf{79}, 677 (2007); R. Saito, G. Dresselhaus, and M.~S. Dresselhaus,
\emph{Physical Properties of Carbon Nanotubes} (Imperial College Press,
London, 1998).

\bibitem {Coiffic2007}J. C. Coiffic, M. Fayolle, S. Maitrejean, L. E. F. Foa
Torres, and H. Le Poche, Appl. Phys. Lett. \textbf{79}, 252107 (2007).

\bibitem {Liang2001}W. Liang, M. Bockrath, D. Bozovic, J. H. Hafner, M.
Tinkham, and H. Park, Nature \textbf{411}, 665 (2001).

\bibitem {Wu2007}F. Wu, P. Queipo, A. Nasibulin, T. Tsuneta, T. H. Wang, E.
Kauppinen, and P. J. Hakonen, Phys. Rev. Lett. \textbf{99}, 156803 (2007); L.
G. Herrmann, T. Delattre, P. Morfin, J.-M. Berroir, B. Placais, D. C. Glattli,
and T. Kontos, \textit{ibid.} \textbf{99}, 156804 (2007); Na Young Kim, P.
Recher, W. D. Oliver, Y. Yamamoto, J. Kong, and H. Dai, \textit{ibid.}
\textbf{99}, 036802 (2007).

\bibitem {Kim2004}J. Kim, H. So, N. Kim, J. Kim, K. Kang, Phys. Rev. B
\textbf{70}, 153402 (2004); Z. Yu and P. Burke, Nano Lett. \textbf{5}, 1403
(2005); C. Meyer, J. Elzerman, and L. Kouwenhoven, Nano Lett. \textbf{7}, 295
(2007); P. J. Leek, M. R. Buitelaar, V. I. Talyanskii, C. G. Smith, D.
Anderson, G. A. C. Jones, J. Wei, and D. H. Cobden, Phys. Rev. Lett.
\textbf{95}, 256802 (2005).

\bibitem {Roland2000}C. Roland, M. Buongiorno~Nardelli, J. Wang, and H. Guo,
Phys. Rev. Lett. \textbf{ 84}, 2921 (2000).

\bibitem {Orellana2007}P. A. Orellana and M. Pacheco, Phys. Rev. B
\textbf{75}, 115427 (2007).

\bibitem {Wuerstle2007}C. Wuerstle, J. Ebbecke, M.~E. Regler, and A. Wixforth,
New J. of Phys. \textbf{9}, 73 (2007).

\bibitem {Guigou2007}M. Guigou, A. Popoff, T. Martin, and A. Crepieux, Phys.
Rev. B \textbf{76}, 045104 (2007).

\bibitem {Oroszlany2009}L. Oroszlany, V. Zolyomi, and C. J. Lambert,
arXiv:0902.0753 (unpublished).

\bibitem {Blanter2000}Ya. M. Blanter and M. B\"{u}ttiker, Phys. Rep.
\textbf{336}, 1 (2000).

\bibitem {Pastawski1992}H. M. Pastawski, Phys. Rev. B \textbf{46}, 4053 (1992).

\bibitem {Jauho1994}A.-P. Jauho, N. S. Wingreen, and Y. Meir, Phys. Rev. B
\textbf{50}, 5528 (1994); C.~A. Stafford and N.~S. Wingreen, Phys. Rev. Lett.
\textbf{76}, 1916 (1996).

\bibitem {Arrachea2005b}L. Arrachea, Phys. Rev. B \textbf{72}, 125349 (2005).

\bibitem {Kurth2005}S. Kurth, G. Stefanucci, C. Almbladh, A. Rubio, and E. K.
U. Gross, Phys. Rev. B \textbf{72}, 035308 (2005).

\bibitem {Agarwal2007}A. Agarwal and D. Sen, J. Phys. Cond. Matt. \textbf{19},
046205 (2007).

\bibitem {Moskalets2002}M. Moskalets and M. B\"{u}ttiker, Phys. Rev. B
\textbf{66}, 205320 (2002).

\bibitem {Camalet2003}S. Camalet, J. Lehmann, S. Kohler, and P. H\"{a}nggi,
Phys. Rev. Lett. \textbf{90}, 210602 (2003).

\bibitem {FoaTorres2005}L. E. F. Foa Torres, Phys. Rev. B \textbf{72}, 245339 (2005).

\bibitem {Keay1995}B. J. Keay, S. Zeuner, S. J. Allen, Jr., K. D. Maranowski,
A. C. Gossard, U. Bhattacharya, and M. J. W. Rodwell, Phys. Rev. Lett.
\textbf{75}, 4102 (1995).

\bibitem {Tien-Gordon}P. K. Tien and J. R. Gordon, Phys. Rev. \textbf{129},
647 (1963).

\bibitem {Shi2009}L. Shi, J. Zhou, P. Kim, A. Bachtold, A. Majumdar, and P. L.
McEuen, arXiv:0904.3284v1 (unpublished).

\bibitem {Strass2005}M. Strass, P. H\"{a}nggi, and S. Kohler, Phys. Rev. Lett.
\textbf{95}, 130601 (2005).

\bibitem {Moskalets2008}M. Moskalets and M. B\"{u}ttiker, Phys. Rev. B
\textbf{78}, 035301 (2008).
\end{thebibliography}


\end{document}